# GALAXYSEARCH –DISCOVERING THE KNOWLEDGE OF MANY BY USING WIKIPEDIA AS A META-SEARCHINDEX

Hauke Fuehres
University of Cologne
Pohligstr. 1
50969 Cologne, Germany
fuehres@wim.uni-koeln.de

Peter A. Gloor
MIT CCI
Cambridge, MA 02139
+1 617 253 7018
pgloor@mit.edu

Michael Henninger
U. App.Sci. NW Switzerland
i4Ds
CH-5210 Windisch
michael.henninger@fhnw.ch

Reto Kleeb
MIT CCI
Cambridge MA 02139
rkleeb@mit.edu

Keiichi Nemoto
Fuji Xerox Co., Ltd.
Tokyo, Japan
keiichi.nemoto@fujixerox.co.jp

## ABSTRACT

We propose a dynamic map of knowledge generated from Wikipedia pages and the Web URLs contained therein. GalaxySearch provides answers to the questions we don't know how to ask, by constructing a semantic network of the most relevant pages in Wikipedia related to a search term. This search graph is constructed based on the Wikipedia bidirectional link structure, the most recent edits on the pages, the importance of the page, and the article quality; search results are then ranked by the centrality of their network position. GalaxySearch provides the results in three related ways: (1) WikiSearch - identifying the most prominent Wikipedia pages and Weblinks for a chosen topic, (2) WikiMap - creating a visual temporal map of the changes in the semantic network generated by the search results over the lifetime of the returned Wikipedia articles, and (3) WikiPulse - finding the most recent and most relevant changes and updates about a topic.

## INTRODUCTION

Frequently the main problem in Web search is not finding the right answer, but asking the right question. This paper introduces a novel information discovery system based on Wikipedia. Wikipedia editors make sure to keep their articles continuously updated: Natural disasters, political news and scandals are constantly monitored in articles or in links between them. But how do these pages and their links evolve over time? Wikipedia does not only provide the digital world with a vast amount of high quality information, it also opens new opportunities to investigate the processes that lie behind the creation of the content as well as the relations between knowledge domains. The goal of our project is to create a dynamic map of knowledge, visualizing the evolution of links between articles in chosen subject areas. The basic idea of GalaxySearch is to use Wikipedia as an index for different types of search queries. GalaxySearch delivers most relevant search results filtered by three criteria, first, the most relevant Wikpedia and Web pages; second, a dynamic semantic map of knowledge generated from the Wikipedia pages; third, the most recent and news-worthy Wikipedia and Web pages.

## RELATED WORK

Our work draws on three strands of related research: (1) Search optimization through Wikipedia, (2) mapping knowledge by Web mining, and (3) automatically identifying latest news from the Web. To the best of our knowledge there is no system, that combines all three related functions, although there is wide research in all three areas.

The Wikiseek engine (en.wikipedia.org/wiki/Wikiseek) was a commercial search engine available from 2007 to 2008, offering a general search service by indexing Wikipedia pages and the URLs contained in Wikipedia. (Hahn et. al. 2010) describe a system using article quality to improve the built-in Wikipedia search. Wikipedia's sister project Wikinews provides latest news collected from traditional news sources and edited by Wikipedians.

There are many systems visualizing ontologies created from Wikipedia. (Weld et. al. 2008) combine WordNet with Wikipedia infoboxes to construct ontologies. Vispedia (Chan et. al. 2009) is providing a Mashup interface to visualize Wikipedia data in geographic maps, timelines and scatterplots. Folksoviz (Lee at. al. 2008) uses Wikipedia to build semantic graphs of delicious tags. (Holloway et. al. 2007) use Wikipedia categories to construct a full network map of all pages in Wikipedia, using color-coding to visualize different dimension such as last edit time, or most active editors. Viegas, Wattenberg and Dave (2004) have built widely quoted visualizations of Wikipedia editing activity called HistoryFlow and Chromogram.

None of the systems described above, however, combines all three elements of using Wikipedia as a search engine index, displaying the results as a semantic network and showing a temporal evolution of the topics.

# SYSTEM ARCHITECTURE

GalaxySearch is a search and knowledge-mapping engine using Wikipedia as its index. It creates a series of meta-indices ranking the Wikipedia pages and the URLs in Wikipedia by different criteria. At the core of the ranking system is a graph constructed by the most relevant Wikipedia pages and URLs in response to a given query (figure 1). This graph can be constructed by different approaches and is described in the subsequent sections.

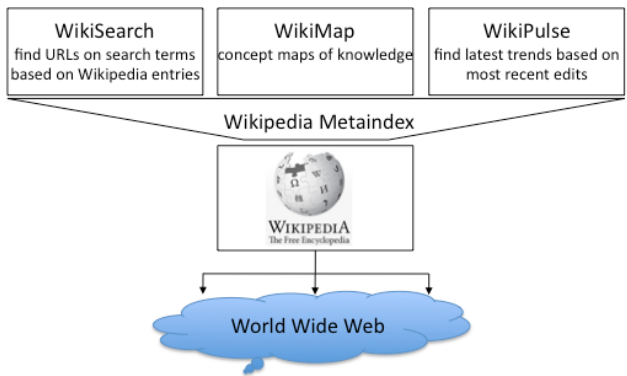

*Figure 1. GalaxySearch Components*

# WIKISEARCH

The core function of GalaxySearch consists of creating a semantic network of the most prominent Wikipedia pages about a chosen topic. These Wikipedia pages are then used to identify the most relevant URLs about the same topic. While the search function provided by Wikipedia offers a good starting point to find the most relevant pages in Wikipedia, it does not return a more fine-grained semantic network of relevant articles that can be used for ranking of the Web pages listed on the Wikipedia pages.

Doing a search for the network of the most linked Wikipedia pages about a subject can be considered "semantic search" because both syntactically similar as well as semantically similar results are returned. This is because outgoing links of a Wikipedia article refer to articles that are similar in content to the referencing article but usually do not include the search words.

We therefore pursue two related goals with Wikisearch: First to get the most relevant semantic network for Wikipedia articles, and second to obtain the most relevant Web pages for a given topic. In Wikisearch we interpret relevance in two dimensions: In return to a search we want to find either "latest news" (for getting news about a the search topic) or "established knowledge" for receiving information grounded in science about a specific topic.

## Gathering the Data

The analysis of Wikipedia articles and their connections is something that has been done extensively in the past (Ganjisaffar et. al. 2009, Nunes at. al. 2008, Yeh et. al. 2009), most of these projects have however worked on static data sets, based on database dumps that are provided by the Wikipedia foundation. The data that is used for this approach usually focuses on the most recent revision of the articles, it has a size of approximately 30 GB (uncompressed, English Wikipedia, 3.6 Million articles) and can be handled with modest hardware requirements.

This dataset does not include any historical data and does not allow the study of changes in the structure of articles over time. In addition to the aforementioned datasets the Wikipedia foundation also provides database files that include the complete historical data, these files however are currently about 5 TB (English Wikipedia, uncompressed). Another factor that reduces the usefulness of these dumps for our requirements to display data as close to real time as possible is that these datasets are only provided about once a month. This makes it impossible to closely monitor the development of events that are currently in progress.

To work around these issues we developed a data fetcher that relies on the Wikipedia HTTP API. The fetcher ensures that the amount of requests to the API is reduced to a minimum. It continuously collects and stores the minimal amount of information that is required to build link-networks for a selected list of articles with the desired timestamp resolution.

The network created by Wikisearch contains two different types of nodes: Web Nodes (external) and Wikipedia nodes (internal). Each Web node is connected to at least one Wikipedia node. This connection represents the link from a Wikipedia article (the node in the graph) to an external URL (the Web node). In the following sections different approaches to rank the retrieved nodes are described.

### Indegree-Counter

Wikipedia offers the possibility to identify all other Wikipedia articles that point to a specific Wikipedia article or URL through the Wikipedia API. This means that we can obtain information on how well connected a node is through one API call.

Unfortunately the indegree counter of a Wikipedia page is not context-specific. The Wikipedia API call returns all referencing articles in the whole Wikipedia. This leads to the problem that very popular topics such as countries, are ranked very highly because they are referenced a lot in other articles. Therefore this approach does not work very well.

### Bidirectional Ranking

A better approach is to look for bidirectional links, computing all the links between a set of Wikipedia pages. The most relevant network is obtained by filtering out all links between Wikipedia articles that are not bidirectional. This reduced network only includes nodes that have a strong semantic connection.

*Quality & Importance Rating*

Quality and importance assessments of Wikipedia articles are performed by members of WikiProjects[1]. They use a 4-point rating range for importance (Low – Mid – High – Top) and a 9-point rating range for quality (List – Stub – Start – C – B –GA – A – FL – FA (highest).[2] In our work we use the "Wikipedia Release Version Tool"[3] that offers easy access to quality and importance measures for Wikipedia articles.

The rating of the Wikipedia search results corresponds to their quality or importance rating. Unrated articles get a zero quality or importance rating. The value of an external URL is calculated by summing up all quality or importance ratings of the referencing Wikipedia articles.

The main disadvantage of this method is that not all Wikipedia articles are rated by importance. Therefore, applying this algorithm leads to articles having low importance rating even if they might be well suited for the given context. On the other hand, the available ratings are very reliable due to the peer review of the ratings by the community members.

*Actuality Rating*

The full edit history of each Wikipedia article is readily available, offering information about the revisions of the articles. As a first crude step towards identifying the most pertinent news in the Wikipedia – we call it "actuality" – we took the numbers of revisions from the edit history during the last two weeks, based on the idea that the more an article changes, the more actual it is. The actuality rating of Web pages is calculated by summing up the number of revisions of all referencing Wikipedia articles.

It turns out that this simple approach for measuring the actuality does not work very well. Many popular articles (e.g. country articles) have a lot of revisions simply because many people feel the urge to add information. For highly specialized topics it is much harder to find experts, thus leading to a much smaller pool of potential editors. This leads to a high actuality rating for articles that are not really "actual" but rather "popular". We are therefore currently integrating a decaying edit history analysis that is outlined in the "future work" section below.

**Finding the Best Search Ranking**

Each of the described algorithms can be used to create a semantic network, using the generated association (bidirectionality, actuality, importance, quality) as an edge in the network. Generating the best network is as much "art" as "science". The first part of the "art" consists of defining the appropriate threshold value for the node rank. All nodes with rating values below this threshold will be removed. The second part of the "art" consists of identifying how to best combine the different types of link networks: bidirectionality, actuality, quality, and importance. The third part of the "art" consists in the way the networks are constructed. In the current version of the galaxySearch system, each of the 4 algorithms gets similar link weights. Nevertheless, to obtain optimal results from the search and visualization perspective, we experimented with different weights for each of the four link types, resulting in vastly different networks for every rating approach and each threshold value.

To calculate and visualize the graph we use the Condor dynamic social network analysis tool (formerly called TeCFlow) (Gloor & Zhao 2004). It allows us to flexibly combine different graphs and calculate different graph metrics. In particular, we experimented with degree and betweenness centrality metrics (Wassermann & Faust, 1994), where a node that occurs many times on the shortest path between other nodes gets a higher value. Usually the originally selected Wikipedia articles for starting the collecting process have high centrality values. A second reason for high centrality might be that the article has many back-links.

This rating delivers good results, especially for Wikipedia articles. Ranking Web links it is somewhat harder because many Web links appear only on one Wikipedia article. Therefore they represent a single leaf in the graph and will end up with a degree of 1 and betweenness of 0, making selective filtering difficult.

See figures 2, 3, 4, and 6 for an illustration. Figure 2 shows the graph of Wikipedia nodes generated by bidirectional links, in response to the query "abortion", with only the "abortion" article chosen as the seed for the query, which is shown in figure 5. Figure 3 shows the graph of the Web URLs collected with the same settings, connected by the Wikipedia articles on which the URLs are listed. Figure 4 shows the full graph of all Wikipedia pages – including the ones with no URLs – and Web URLs. Figure 6 shows the semantically linked network of Wikipedia pages returned in response to the query "Dominique Strauss-Kahn", generated by combining all four raking algorithms, with only the top 50 article names by betweenness shown. Note the blue cluster of celebrities at the top of the figure, created through the actuality ranking. It owes its existence to the Wikipedia page "Time_100" about the 100 most influential people as assembled by Time Magazine. Dominique Strauss Kahn made it on this list in 2010, which lead to inclusion of this page. Once it was there, the many edits on the different celebrity Wikipedia pages guaranteed high ranking by the actuality algorithm, although the semantic link of many celebrities to Dominique Strauss Kahn seems quite strenuous.

**Measuring Search Performance**

To evaluate the performance of the different Wikisearch ranking algorithms through precision and recall, we use the normalized discounted cumulative gain evaluation method at top k nDCG@k (Järvelin & Kekäläinen, 2002) which is widely used to measure success of Web searching (Hahn et. al. 2010).

---

[1] http://en.wikipedia.org/ wiki/Wikipedia:WikiProject
[2] Assessment of Wikipedia articles: http://en.wikipedia.org/wiki/Wikipedia:Version_1.0_Editorial_Team/Assessment
[3] http://toolserver.org/~enwp10/

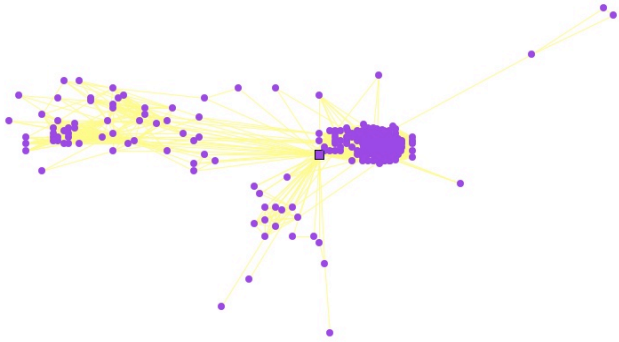

*Figure 2. Network of Wikipedia articles only about "Abortion", nodes ranked by bidirectionality*

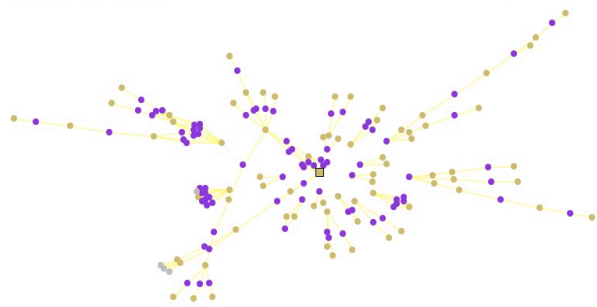

*Figure 3. Network of external URLs in Wikipedia articles about "Abortion", linked by only the Wikipedia articles where they are mentioned, nodes ranked by bidirectionality. Pink nodes are Wikipedia articles, brown nodes are URLs.*

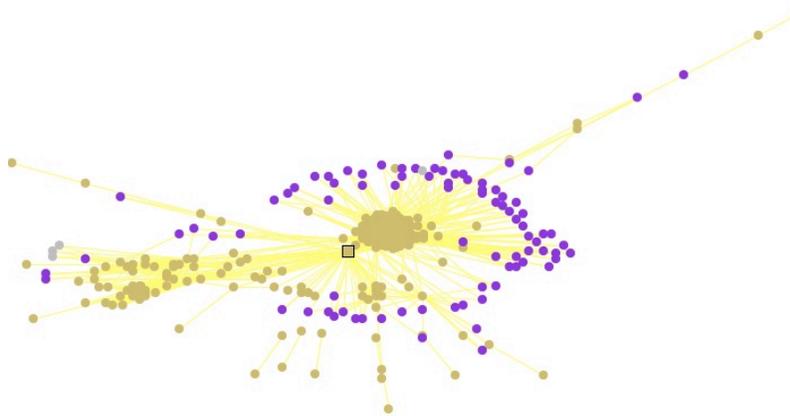

*Figure 4. Full combined network: Wikipedia Articles from figure 3, and external URLs from figure 4, nodes ranked by bidirectionality. Pink nodes are Wikipedia articles, brown nodes are URLs.*

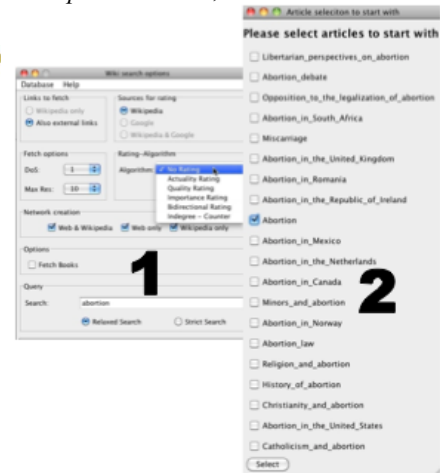

*Figure 5. Wikisearch user interface. 1: choose ranking algorithm, 2: choose starting pages for collection, list is generated by pages returned by Wikipedia's built in search function*

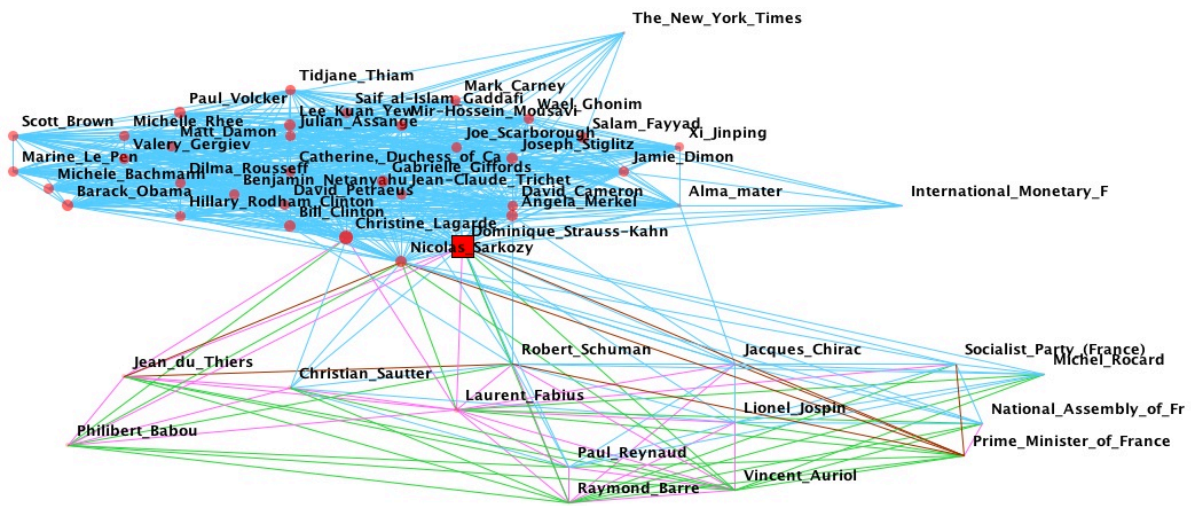

*Figure 6. Semantic Network (wiki-links only) about "Dominique Strauss-Kahn" bidirectional(pink), importance(brown), quality(green), actuality(blue) links combined, only top 50 nodes by betweenness shown*

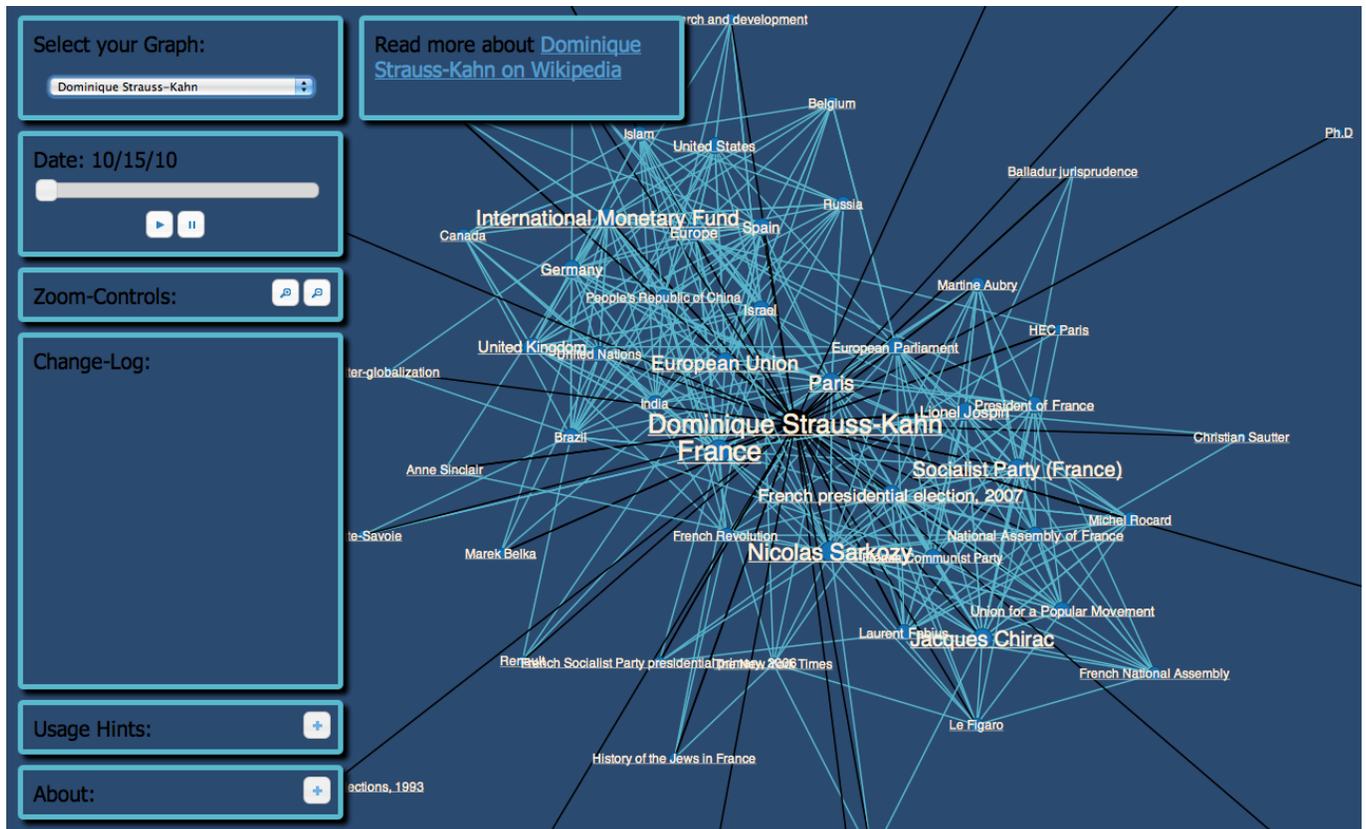

*Figure 8. Wikimaps Graph of Dominique Strauss-Kahn on October 15th 2010 (online at http://www.ickn.org/wikimaps)*

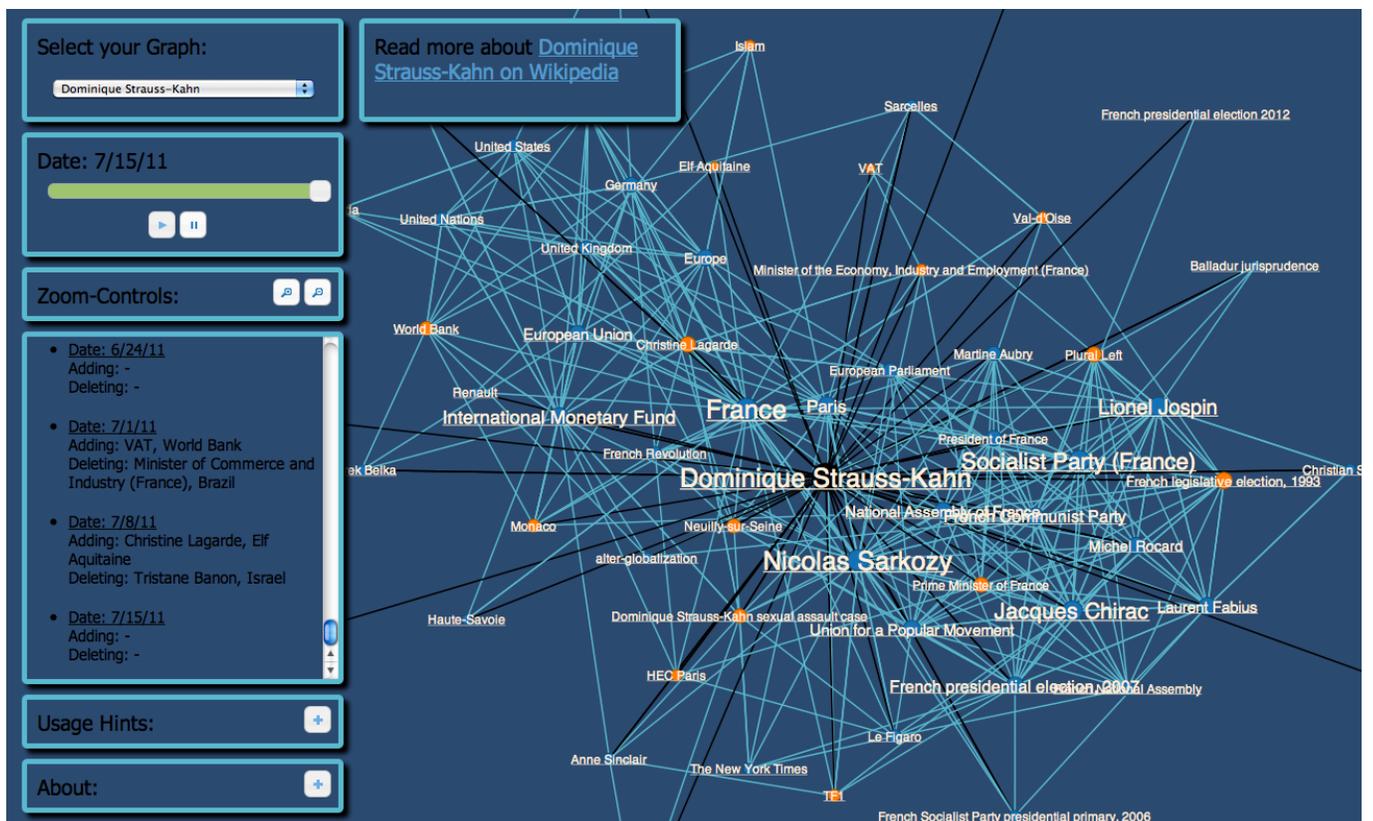

*Figure 9. Wikimaps Graph of Dominique Strauss-Kahn on July 15th 2011*

$$s(p) = \begin{cases} 2 & \text{if } 1 \le p \le n_q^{HR} \\ 1 & \text{if } (n_q^{HR}+1) \le p \le (n_q^{HR}+n_q^R) \\ 0 & \text{otherwise} \end{cases}$$

$n_q^{HR}$ is defined as the number of articles within the k search results that are of high relevance (HR) (rating = 2).

$n_q^R$ is defined as the number of articles within the k search results that are relevant (rating = 1)

The gain G for query q for the top k search hits is then computed as follows:

$$G_q(k) = \frac{1}{N_q} \sum_{p=1}^{k} \frac{2^{r(p)}-1}{\log(1+p)}$$

r(p) is the reward giving to a search hit at position p based on whether it is of type HR (r(p)=2), of type R (r(p)=1, or not ranked (r(p)=0). N is a normalization factor for query q, making sure that a perfect ordering of articles would lead to G(k)=1.

$$N_q = \sum_{p=1}^{k} \frac{2^{s(p)}-1}{\log(1+p)}$$

We measured nDCG@k for the two types of search "established knowledge" and "latest news" introduced before. In the following discussion we illustrate our results by looking at the query for "abortion" ("established knowledge") and the queries "Dominique Strauss Kahn" ("latest events") and Syria (a combination of "latest events" and "established knowledge").

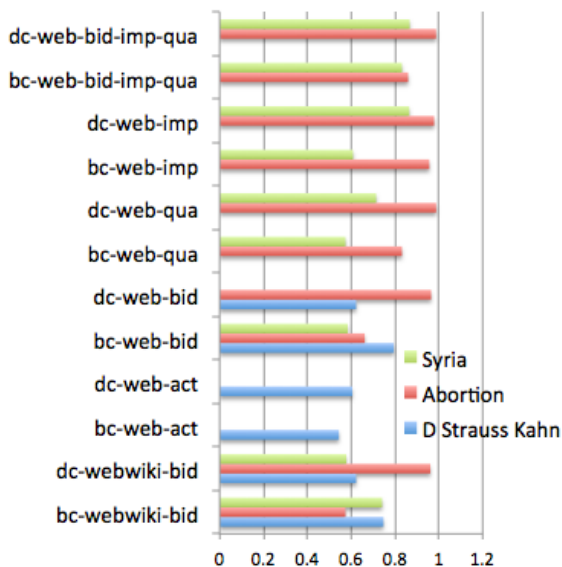

Figure 7. nDCG@k of different search algorithm combinations (dc:degree centrality, bc: betweenness centrality, imp:importance, qua:quality, bid:bidirectional, web: web-only network (fig 3), webwiki: full network (fig 4)

As the results in figure 7 illustrate, search quality differs widely depending on the algorithms employed and the type of query "latest news" or "established knowledge". In general, "established knowledge" queries lead to higher quality results than "latest news" queries.

For "established knowledge" good results are provided by each rating, and by combining multiple ratings, taking degree centrality of each node in the graph as the ranking metric. Using the combined Web-Wikipedia article graph (fig. 4) instead of the Web graph (fig. 3) does not lead to better results.

For "latest news" queries results are very different. They are in general worse, and they can be improved by creating graphs combining multiple ranking algorithms. Figure 6 illustrates potential distortions by "hijacking" of queries through celebrity pages, in this case the links to the Times 2010 "100 most influential people" list. In the future work section we discuss potential remedies to this problem.

## WIKIMAP

WikiMap consists of two separate parts: the first part is the data gathering- and filtering component based on WikiSearch that collects and caches all the relevant graph information. The second part is does the rendering of the animation of the graph's evolution over time in a Web browser, showing it in a more accessible and visually appealing way than the desktop based Condor.

### Generating all Graphs for the Timeline

The algorithm to create a meaningful network of nodes for any search term is divided in two steps: the first step includes the collection of all nodes that potentially could be included and the creation of a graph containing all potential candidates. The second step consists of the reduction of the number of nodes by applying filters that are based on information collected using the graph.

Before the filtering step the graph typically contains between 1000 and 2500 articles. The applied filters then drastically reduce this number, based on the number of connections to the original node ordered by indegree and shortest path, leading to a graph with a more manageable number of nodes of less than 50.

The final graph is visualized using the Protovis[4] framework that is based on JavaScript and SVG. It provides the algorithms to calculate the position of the nodes based on a force directed layout (Fruchterman & Reingold 1991). Protovis further computes the radius of the nodes proportional to the indegree. The indegree then affects the virtual electrical charge in the physical model that defines the layout, resulting in a more central position of nodes with a high indegree.

### Tracking Changes in the Graph over Time

Once the initial set of articles is collected, the filtering algorithm can be repeatedly run for any given array of time stamps. The resulting set of graphs can then be incorporated into an animation that allows users to track the evolution of a given topic network over time. The chosen algorithm for the layout makes it easy to visually detect subject clusters and allows a visual assessment of the strengthening or weakening of ties between clusters (Gloor et. al. 2004). See figures 8 and 9 for an example.

---

[4] http://www.protovis.org

This way a combination of graph properties can successfully be used to gather high-relevance results within Wikipedia. The visualization of these results allows users to get a simple overview of related topics and the evolution of these topics over time. An alpha version of WikiMap is available online at http://www.ickn.org/wikimaps.

## RESTRICTIONS AND FUTURE WORK

While our approach works well for finding relevant articles and Web URLs on Wikipedia for "established knowledge", it does not work so well for "latest news". While starting with the bidirectional link network and subsequently adding other types of networks seems to be a promising approach, further work is needed to improve the actuality component of our system. For "latest news" we are therefore trying a different approach.

Our goal is to create "WikiPulse", a system dedicated to finding and visualizing latest news from Wikipedia. It builds on WikiSearch and WikiMaps, customizing the search function for actuality. The goal of WikiPulse it to discover current events while they are being discussed and described in Wikipedia. We measure success by comparing our results to WikiNews, Google News, and Bloomberg. As (Elsas & Dumais, 2010) have found, Web pages of higher relevance change more than pages that are less relevant. This means that searching for pages with many edits should be an additional predictor of high relevance. However, as we already discussed, employing the straightforward approach only delivers mediocre results.

One idea is to count the number of edits during the last two weeks for the user-selected articles in the beginning, analyzing the revisions and identifying which parts of the article changed. Then not all links from the articles are collected but only those within the changed sections. Multiple changes in the same section can be represented by link weights. This can also be extended to "latest news during a specific time period". This way a user will be able to study what was popular about a given topic at a specific point of time in the past.

Another area for improvement is the temporal backwards searching as employed in WikiMap. The initial search step of the algorithm is based on the most recent data, because the Wikipedia API does not offer historical search results. To get accurate data for the backwards-searching step we will need to create a system that stores searchable indexes of the complete Wikipedia dataset for any desired timestamp.

Besides the obvious network structure of Wikipedia articles, there is a second source of data than can be further incorporated to improve the quality of search results in WikiMaps, namely the "shared-editorship" network of Wikipedia authors, where a link between two articles is made if the two articles have been edited by the same author. First results creating semantic networks using this approach are encouraging (Nemoto 2010).

Another area for improvement, which we have not tapped yet is making use of the categorization of the articles. It has been shown by (Vercoustre et al. 2008) that the categorization of articles in Wikipedia can be harnessed to further improve the accuracy and relevance of search results.

While this project is in an early stage, it builds on three years of research in our group, studying Wikipedia co-authorship and edit-networks (Nemoto et. al. 2011), as well as a vast body of research in Wikipedia authorship and content by a vibrant global research community. We already have been able to show that Wikipedians form long-lasting collaboration network resulting in high quality output. We are convinced that including these and other results will help us in building a new lens into the knowledge of mankind captured in Wikipedia, providing – we hope – yet another stepping stone towards more creativity and innovation.